# TD-MQTT: Transparent Distributed MQTT Brokers for Horizontal IoT Applications

Fatma Hmissi
*CRISTAL Labotory, ENSI*
*Manouba University Campus*
Manouba, Tunisia
Email: fatma.hmissi@ensi-uma.tn

Sofiane Ouni
*CRISTAL Labotory, ENSI*
*Manouba University Campus*
Manouba, Tunisia
Email: sofiane.ouni@insat.rnu.tn

*Abstract*—MQTT (Message Queuing Telemetry Transport) has become the perfect messaging protocol for IoT (Internet of Things) systems since it is the lightest protocol designed for lowbandwidth, high-latency, unreliable networks. Today, the strategy of distributing several MQTT brokers on the networks is widely used because the strategy of using a single broker is no longer efficient. However, in the distributing architectures of MQTT brokers, a subscriber should have prior knowledge about the address of the broker that publishes the data on the topics of interest. In this paper, we tackle this challenge by proposing a mechanism that connects the subscribers to the brokers in a transparent way. The proposed approach, known as TD-MQTT (Transparent Distributed MQTT brokers), requires no prior knowledge of the brokers by the subscribers. The data will be carried automatically from brokers that can change their configuration and location. The transparency will help to use IoT data without worrying about their location and dynamic configuration changes. To evaluate our approach, we compared it with the basic distributed MQTT and the EMMA (MQTT Middle-ware for Edge Computing Applications) approach. The results of the evaluation show that TD-MQTT is much better than the standard MQTT, especially in terms of response time.

## I. INTRODUCTION

Since its inception in 2000, the Internet of Things (IoT) has grown in popularity [1]–[3], where the number of connected devices predicted to reach 75 billion by 2025 [4]. The IoT allows devices to communicate among each other. For that purpose, several communication protocols were proposed. These protocols are classified into three categories, namely, messaging protocols, device management protocols, and service discovery protocols [5]. Nowadays, several messaging protocols have been implemented to tackle the problem of the limited resources of IoT devices, e. g. , memory and battery, where the messaging protocol Message Queuing Telemetry Transport (MQTT) [6], [7] is the most widely used in IoT projects. This choice is due to the lightness and flexibility of the MQTT protocol [5]. IoT projects based on basic MQTT communication have a centralized architecture where an MQTT broker is hosted on a public or private Cloud. All publishers produce the data for that broker and the subscribers' interest in a topic in the same broker. This reasoning shows a performance degradation in IoT systems.

Hence, sending a large amount of data to the broker results network congestion and increases the delay of response. Fog, Edge and Mist Computing architectures are the solutions to Cloud Computing issues. In these architectures, several MQTT brokers are distributed on Fog, Edge and Mist levels. Hence, several MQTT brokers are distributed in network to collect and forward the data.

In IoT systems based on the distributed MQTT architecture, several MQTT brokers are deployed at the Mist, Edge, or Fog layer of the network. In the typical IoT communication based on the distributed MQTT publish-subscribe architecture, when IoT devices generate new data, they update the given topics by publishing PUBLISH messages to brokers. Once the brokers receive these messages, they update the given topics and consequently broadcast the new data to all the clients subscribed to these topics.

Now, we briefly summarize the main scenario that motivates the need for a transparent distributed MQTT broker approach. Currently, the market for IoT applications is subdivided into two categories: vertical and horizontal [3], [6]. A vertical IoT application is software that serves the demands of a single industry and is given and managed by the same organization, whereas a horizontal IoT application links several users with diverse knowledge and skill sets by enabling multiple suppliers to work with a shared framework. In horizontal applications, new subscribers may be added to the system continuously, and so on, brokers by several industrials. But, each industrialist should know the addresses of the brokers that contain the topics of interest. This strategy seems to be the major problem with horizontal applications. Since , users should be aware of framework updates at all times. This situation requires prior knowledge of the MQTT brokers' addresses, specifically, every subscriber should know the location of the broker who is managing the topic of interest before joining the network or, should connect to a broker that does not manage the topics of interest, and consequently, that broker should communicate with the other brokers on the network to retrieve the topics of interest. A number of questions arise when designing such communication. (i) When a subscriber loses connectivity with a broker due to a broker malfunction or a loss of Internet connection, the subscriber should be assigned to another broker. (ii) If a subscriber does not find any more of the topics of interest at that broker, he should be associated with the broker who manages these topics. Consequently, we designed a new approach to transparently connect the subscribers to the distributed MQTT brokers, called, transparent distributed MQTT brokers (TD-MQTT). The designed approach allows the subscriber to get the broker's address at the time of joining the network and periodically update this address according to the location of the topics of interest and the broker's status.

The remainder of this paper is organized as follows, section II gives a briefly overview of the MQTT protocol and the features that will be used in TD-MQTT. Section III overviews the related works. Section IV introduces and describes TDMQTT and the innovative features of this approach. Section V presents the performance evaluation of the proposed approach. Section VI concludes the paper.

## II. BACKGROUND

The abbreviation MQTT stands for "Message Queuing Telemetry Transport". MQTT was designed to be used in Machine-to-Machine (M2M) and Internet of Things (IoT) systems for messaging. MQTT is characterized by its lightweight and flexibility, making it perfect for IoT communications. This protocol belongs to the application layer of the Open Systems Interconnection (OSI) model. MQTT's architecture is based on the publish-subscribe model.

The main components of the MQTT's architecture are MQTT clients and the broker. MQTT clients are classified as either MQTT publishers or MQTT subscribers. A publisher is an MQTT client that produces the data, a subscriber is the one who receives it, and a MQTT broker is the server that backs up the data that has been published and matches the data to the appropriate subscriber. There are presently two approved OASIS (Organization for the Advancement of Structured Information Standards) standards of MQTT protocol: MQTT 3 [8] and MQTT 5 [9]. In effect, MQTT 5 has made significant changes to MQTT 3 without increasing overhead or decreasing ease of use. MQTT 5 [9] supports new features such as shared subscription, payload format, content type, and reason codes for all confirmation packets. MQTT 5 [9] and MQTT 3 [8] OASIS standards provide several MQTT control packets to ensure the communication between subscribers and publishers. The MQTT control packets used in TD-MQTT are highlighted in the table I. The OASIS standard of MQTT 5 [9] allows subscriptions to be subscribed to at multiple levels within a topic. For that purpose, the wildcard character # is used. This character must be specified either on its own or following a topic-level separator.

## III. RELATED WORKS

In this section, we briefly discuss related research efforts and explain why we focus on transparent, distributed MQTT brokers.

Seen the multiple advantages of the distributed architectures of MQTT brokers, e.g. , reducing network congestion and guaranteeing limited response time, the use of these architectures has seen a significant rise. Today, many researchers are based on the use of distributed architectures of MQTT brokers in their research papers, namely the works in [10]– [15]. However, only very limited attention has been given to the problem of transparent connectivity to distributed MQTT brokers. Thus, an MQTT client, particularly the subscriber, should have prior-knowledge of the brokers' addresses where the topics of interest reside. Some efforts, such as the work [16], [17], propose a transparent, distributed architecture of MQTT brokers. Now we'll go over the key contribution of each research study in more detail.

J. H. Park et al. [10] encourage the usage of MQTT brokers distribution at the edge level to tackle the problem of massive communication in an IoT architecture based on Cloud Computing. The goal of this work is to reduce data transmission delays. To that end, J. H. Park et al. [10] propose constructing per-topic multi-cast groups, in which they collect information about clients and their relevant pub/sub topics using a Software Defined Networking (SDN) controller.

E. Longo et al. [11] also suggested the use of the distributed architecture where distributed MQTT brokers are moved to the Edge of the Network. E. Longo et al. [11] proposed a new extension of the MQTT protocol, known as MQTT-ST [11] (MQTT-Spanning Tree). MQTT-ST [11] has been developed to address the problem of message loops among brokers in the distributed architecture of MQTT brokers. MQTT-ST [11] is a protocol that links MQTT brokers in a tree topology automatically. Spanning Tree Protocol (STP) and Mosquitto bridging are used in MQTT-ST.

L. Stagliano et al. [12] argue about the importance of the` distributed version of the MQTT protocol. In this version, MQTT brokers connect among themselves. L.Stagliano et` al. [12] proposed a new protocol called Distributed-MQTT, which is a distributed version of the MQTT protocol. DMQTT improves the MQTT protocol by adding new features. The new features include automated broker discovery, an optimized overlay network, and efficient publication routing. The broker discovery feature allows the MQTT brokers to discover each other rather than have prior knowledge of all brokers. Precisely, a broker requires no prior knowledge of the addresses and ports of all other active brokers. For this purpose, L. Stagliano et al. [12] use an IP (Internet Protocol)` multicast group, where each broker sends a discovery message to the IP multicast group on a regular basis, announcing its own IP address. For a pre-defined amount of time, each broker listens for incoming discovery messages from other brokers to know the addresses of all active brokers. And for the optimized overlay network feature, L.Stagliano et al. [12] used MQTT-ST` [11]. Finally, they proposed a new routing protocol that allows efficient publication routing, where they proposed forwarding published messages only to brokers who are interested in such messages.

As we noticed before, the distributed MQTT brokers should cooperate with each other. In particular, a subscriber should have prior knowledge of the broker to which it will be connected. However, the latter may be interesting for topics that do not exist at that broker. Hence, that broker should

MQTT broker is hosted at a public address (a CloudMQTT broker was used). A JavaScript Object Notation (JSON) file is used to store information about every MQTT device on the network. The "Resource Collection" is the name of this JSON file. New devices may be registered, and existing devices may be moved, therefore the Resource Collection is periodically updated.

Additionally, T. Rausch [16] provides a distributed broker network by utilizing an external agent (controller) to monitor each broker's status. The proposal's main goal is to connect many clients to accessible brokers while taking into account dynamic broker network topology and resource availability retrieve these topics from the other active broker. In that way, MQTT brokers should communicate with each other. For that end, new contributions were proposed, including the works in [13], [14].

TABLE I MQTT CONTROL PACKETS USED IN TD-MQTT

| Name | Flow's Direction | Description |
| --- | --- | --- |
| CONNECT | Client to Broker | Request for a connection |
| CONNACK | Broker to Client | Connect acknowledgment |
| SUBSCRIBE | Client to Broker | Request to register the client's interest to topics, the payload of the SUBSCRIBE request contains topics' names. |
| SUBACK | Broker to Client | Subscribe acknowledgment, which validates that a SUBSCRIBE packet has been received and processed. The names of the topics are not included in the header or payload of the packet. |
| PUBLISH | Client to Broker<br>Broker to Client | Transport application messages. The payloads of these packets contain the published data, while the headers contain the names of the topics on which the data is published. |
| PUBACK | Client to Broker<br>Broker to Client | Publish acknowledgment, which confirms that a PUBLISH packet has been received. |
| DISCONNECT | Broker to Client | This indicates that the topic filter is correctly formed but is not accepted by this broker to indicate this the reason code 0x8F is used . To that end, the broker must insert the reason code 0x8F in the field of the Reason Code of the DISCONNECT packet. Also, the broker may inform the client to temporarily change or permanently change its broker with a DISCONNECT packet containing the Reason Code 0x9C (temporarily change) and 0x9D (permanently change). |
| DISCONNECT | Client to Broker | Close the connection normally, the field Reason Code in this case contains the code 0x00. |

E. Pereira et al. [15] proposed a new version of MQTT known by MQTT-RD. MQTT-RD is for MQTT-Resource Discovery. MQTT-RD is a distributed resource discovery architecture based on MQTT protocol for M2M communications. In the MQTT-RD architecture, there are devices responsible for the resource discovery. Each device represents a sniffer that is responsible for discovering resource. There are two types of sniffer in MQTT-RD architecture: Local sniffer and Internet sniffer. Local sniffer is used to represent a Local Area Network (LAN), in a LAN there is at least one local sniffer. Internet sniffer is used to connect with other Internet Sniffers in other networks using the Internet. Every single sniffer is associated with an MQTT broker to allow communications between components. Internet Sniffer uses an additionally MQTT broker for communication over the internet. For this purpose this variations. Publishers and subscribers are linked to available brokers via gateways, which can be implemented for local or remote networks. Hence, the connection between the MQTT clients (publishers and subscribers) and the brokers is transparent since the MQTT clients do not known the address of the MQTT brokers (they communicate between each others using the gateway). The proposed approach is known by EMMA.

Rather of utilizing the basic MQTT protocol, which connects publishers and subscribers to a single broker, R.Kawaguchi et al. [17] propose using distributed MQTT, which uses many brokers to raise the number of publishers and subscribers in the network. R.Kawaguchi [17] implement a transparent distributed broker network that connects MQTT clients to distributed brokers via a gateway. The gateway agent acts as a link between MQTT clients and MQTT brokers, connecting them via the topic name included in published or subscribed messages. This method is used to reduce broker load and support a system's heterogeneous brokers.

## IV. Proposed Approach

This section presents a new extension of the MQTT protocol that allows transparent connections to the distributed MQTT brokers. We call the proposed approach Transparent Distributed MQTT Brokers (TD-MQTT). Next, we describe TD-MQTT and its innovative features, namely, resource discovery, transparent topic subscriptions, and resource anomaly detection.

### A. Approach Description

The proposed approach is based on distributed publishsubscribe architecture where the publishers and subscribers use the MQTT protocol for communications. This approach consists of four components. These components are 1) publishers, which are IoT devices that produce data; 2) subscribers, which are IoT devices that receive the published data;3) MQTT brokers, which are in charge of managing topics and messages, and 4) an MQTT master broker, which is in charge of resource discovery and managing the transparent connection between subscribers and brokers. This broker is hosted on the Edge or in the Fog layer. One of the main goals of introducing the Master Broker was to tolerate failures in standard brokers.

But an important question may arise about the failures of the master broker. Because of the master broker's unavailability, the communications between the subscribers and the standard brokers are interrupted. The discussed issue leads to the detracting of the IoT systems' performance. As a consequence, the master broker is seen as a single point of failure since, if this one fails, the overall IoT system fails to operate. So, a solution for coping with master broker downtime is needed. To tackle this problem, it is highly recommended to provide several master brokers in the network. Hence, migrating the work of the damaged master broker to another one is the solution. Consequently, the IoT system continues working without any problems.

The fundamental principle of the proposed approach is outlined bellow. Instead of connecting directly to a known broker, the subscriber asks the master broker about the location of the broker that he should be associated with for a given topic at the time of joining the network. Later, the master broker responds to that question by directing the subscriber to the appropriate broker, and consequently, the subscriber connects to that broker to subscribe to the given topic. That procedure is called "transparent topic subscriptions". But before the transparent topic subscription procedure, the master broker should know the available MQTT brokers on the network and the topics that each broker manages. To that end, the master broker performs the resource discovery procedure. A subscriber may lose the connection with the broker or the location of the topic of interest may change. Hence, the proposed approach should handle a procedure for broker detection anomalies to reconfigure the connection between subscribers and brokers accordingly.

We deduce that our proposed approach is built on three key procedures: resource discovery, transparent topic subscriptions, and broker anomaly detection. Following that, we'll go over each operation separately from the others.

### B. Resource Discovery

In our proposed approach, we implement an operation of resource discovery. That procedure is achieved by the MQTT master broker. The aim of this procedure is to find all available brokers on the network and, consequently, to list all topics known by every discovered broker. Hence, to achieve the resource discovery procedure, the master broker should go through two sub-operations, namely, brokers discovery and topics discovery.

*a) Brokers Discovery:* To collect the list of the MQTT brokers on the network. We implemented the operation described in algorithm 1. By implementing this program the master broker will be able to sniff each active MQTT broker in the network. The master broker open a port for listening on MQTT brokers ports. The master broker creates a TCP connection with every single IP address in a specific range of address. Once the master broker ensure that he and the MQTT broker are connected on a that specific port, he adds this address to the list of the available MQTT broker.

---

**Algorithm 1** Broker Discovery Function
{B: a list that contains active broker in the networks}
{$b_i$: contains the information related to broker $i$, such as the IP address. } {AddressRange: a table contains the addresses of all the nodes in the network. }
**Input** : AddressRange
**Output** : B
**function** BROKERDISCOVERY(AddressRange)
  $B \leftarrow \{\}$
  **for** $i \neq length(AddressRange)$ **do**
    $addr \leftarrow AddressRange[i]$
    $TCPConnection(adr, MQTT_{Port})$
    **Wait TimeOut do**
    $response \leftarrow TCPResponse()$
    **if** $response$ **then**
      $b_i.addr \leftarrow addr$
      $B \leftarrow B \cup \{b_i\}$
  **return** $B$

---

*b) Topics Discovery:* After the execution of the brokers discovery procedure, the master broker has a list of all available brokers on the network, where $B = \{b_1, b_2, ..., b_n\}$. Now, it's time to search all topics for each MQTT broker in the list of the discovered brokers. Formally, the master broker looks for $Topics(b_i) = \{topic_1, ..., topic_l\}$ which is the set of topics of each broker $b_i$, where $b_i \in B$ and $i \in \{1..n\}$, n is the number of broker where $n = |B|$. The relevant question to be asked here is how the master broker can list the topics known to an MQTT broker. At this stage, we will respond to this question. For that purpose, we provide the procedure "Topics Discovery". The figure 1 shows the main principle of the "Topics Discovery" procedure by introducing the sequence diagram of the messages exchange between the master broker and the broker $b_i$. First, the master broker should use the CONNECT packet to connect to the target broker ($b_i$), and then the broker ($b_i$) should return a CONNACK message indicating that the connection was successful. After a successful connection, the master broker sends to the given broker a SUBSCRIBE Packet to indicate interest in all the topics using the multi-level wildcard character (#) so the master broker subscribes to "#" to receive every Application Message, and then the given broker returns a SUBACK Packet indicating that the subscriptions were created successfully. The broker $b_i$ sends a PUBLISH packet to the master broker, to forward data where published to topic matching this subscription. Since the variable header of the PUBLISH message contains the field Topic Name where the broker identify the information to which Payload data is published, the master broker retrieve these information to list name of topics known by this broker. After that, the master broker returns a PUBACK Packet indicating that the Application Message was successfully received. And finally, the master broker sends a DISCONNECT Packet to the broker to close the Network Connection.

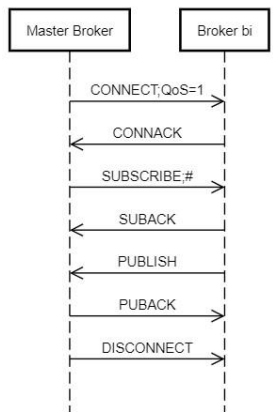

Fig. 1. Topics Discovery Procedure

C. Transparent Topic Subscriptions

Now, we'll go through "Transparent Topics Subscriptions" procedure in more detail. Following the resource discovery procedure, the master broker has two types of lists: $B = \{b_1,..,b_n\}$ that refers to all the MQTT brokers on the network, and $Topics = \{Topics(b_1),..,Topics(b_n)\}$, which refers to the set of all topics known by each MQTT broker, where $Topics(b_i) = \{Topic_1,..,Topic_l\}$ and $l = |Topics(b_i)|$. The figure 2 summarizes the principle of the procedure Transparent Topics Subscriptions. If an IoT device wishes to subscribe to $topic_j$, where $topic_j$ is provided by the broker $b_i$, the subscriber sends a request to the master broker to communicate with him the target broker's address. For that purpose, the subscriber should first of all open a network connection with the master broker by sending to him a CONNECT packet. After that, the master broker will indicate in a CONNACK Packet that the connection is open. Now, the subscriber can send his request for the target broker's address by sending a SUBSCRIBE packet to the master broker indicating the name of the topic that interests him ($Topic_j$). Before sending the target broker's address, the master broker informs the subscriber that his request has been successfully received by sending a SUBACK message. The master broker search now the $Topic_j$ in the set of topics and consequently sends a DISCONNECT Packet to redirect the subscriber to the appropriate broker $b_i$. The DISCONNECT Packet contains a reason code and the broker reference. A reason code is a onebyte unsigned value that indicates the result of an operation. The Reason Code value of 0x9C indicates using another server, while the value of 0x9D indicates that the server has been moved. A broker reference consists of the broker's name, followed by a colon, and the port number. A host name, DNS (Domain Name System) name, SRV (Service Location) name, or literal IP address is the name of a broker reference. Once the subscriber receives a DISCONNECT message from the master broker, he is allowed to use the new broker's location provided and connect to it immediately, and consequently, receive the data published to the topic $Topic_j$.

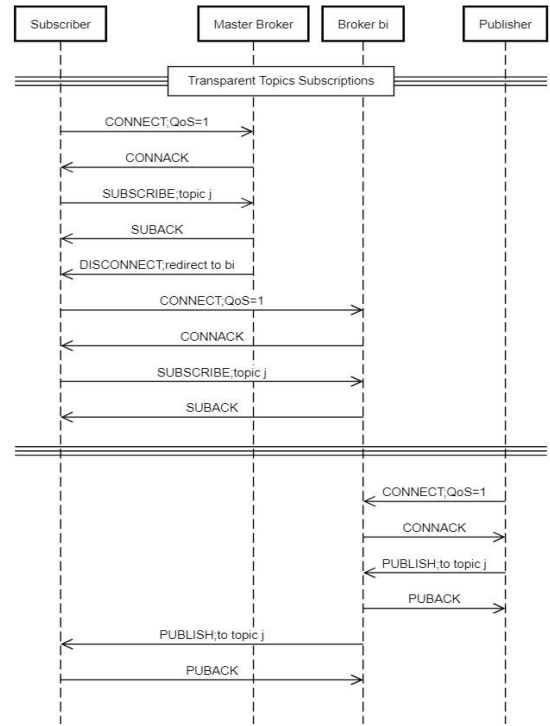

Fig. 2. Transparent Topics Subscriptions

## D. Broker Anomaly Detection

Our approach provides a mechanism for detecting anomalies (e. g, equipment and software failures, loss of connection ) in MQTT brokers so that the connections between the MQTT brokers and the subscribers can be reconfigured appropriately. In this scenario, two methods were used sequentially to examine: 1) the malfunction of the MQTT brokers by exchanging of alive messages and 2) the new topic's location.

To examine the malfunctions of the MQTT brokers, we proposed a method that is included in the subscribers that allows for the exchange of live messages. That message consists of a PINGREQ packet being sent periodically. If the PINGREQ Packet is not received by the MQTT broker $b_i$ in the network within a certain amount of time ($TimeOut$), the subscriber restarts the procedure "Transparent Topics Subscriptions" to request the location of the new broker from the master broker, hence, the master broker update the list of brokers and the topics delivered by each broker and then respond to the demand of the subscriber by sending a DISCONNECT message containing the name of the new broker as shown in the figure 3.

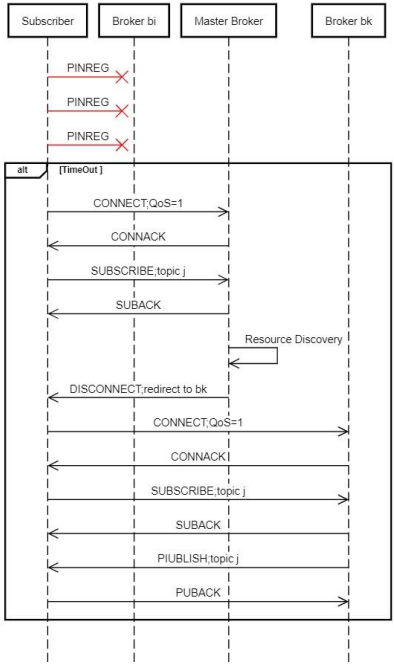

Fig. 3. Broker Anomaly Detection

On the other hand, if the topic is moved from the broker, a method is included in the MQTT broker that allows the exchange of a DISCONNECT packet to inform the subscriber that the topic's location has been changed. If the DISCONNECT packet is received by the subscriber, the latter will restart the operation Transparent Topics Subscriptions, as illustrated in the figure 4, in this level the master broker update the list of the topics of each broker to recover the address of the broker that have the topic of interest before sending the DISCONNECT message.

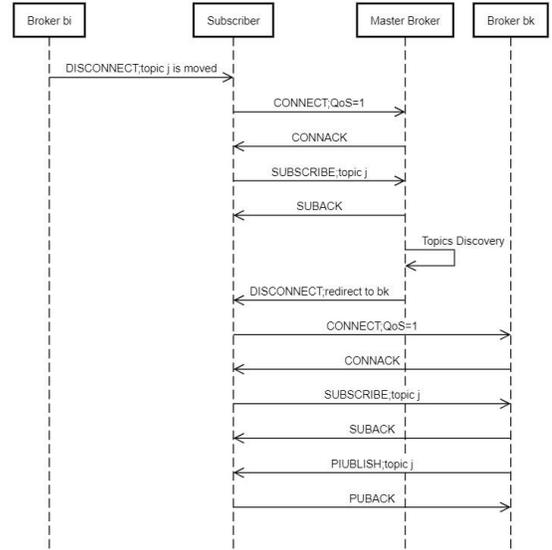

Fig. 4. Topic not Acceptation Detection

## V. Performance Evaluation

This section aims at understanding the performance of the proposed approach (TD-MQTT) by measuring the response delay and the overhead in the data subscription process between subscribers and the broker. The outcomes of the evaluation are compared to those of the standard MQTT and EMMA approaches.

### A. Evaluation Model

To evaluate TD-MQTT, we defined the use of the below equations. To transmit a message over the network we need a delay equals to $T_{message}$. The transmission delay of message equals to size of that message (size) divided by transmission rate of the network (Throughput) as as shown the equation 1.

$$T_{message} = \frac{size}{Throughput} \quad (1)$$

Based on the M/M/1 queuing model described in [18], we considered that an MQTT broker is an M/M/1 system, where a message remains in a broker for an estimated period of time equals to queuing time plus service time (noted $D$). Equation 2 illustrates the time that a message remains in an MQTT broker ($T_{MR}$) [18]. Where $D = \frac{1}{\mu}$, $\mu$ is the service rate of messages and $\lambda$ is the message arrival rate to the broker.

$$T_{MR} = \frac{D}{1 - (\lambda D)} \quad (2)$$

Now a master broker needs a delay $T_{TD}$ to discover the list of the topics available on an active broker. This delay depends on the number of the messages exchanged between the master broker and an active broker. Seven messages are exchanged, namely CONNECT, CONNACK, SUBSCRIBE, SUBACK, PUBLISH, PUBACK and DISCONNECT messages. So $T_{TD}$ equals to the delay needed to transmit all those message as evidenced by equation 3. We used the equation1 to calculate the transmission delay of each message. We noted the delay for transmitting a message $T_{typeofmessage}$ for simplicity; only the first letter of the type of message is used to indicate the type of message, e. g. , $T_C$ for $T_{CONNECT}$.

$$T_{TD} = T_C + T_{CA} + T_S + T_{SA} + T_P + T_{PA} + T_D \quad (3)$$

And for the brokers discovery process, the master brokers needs a time $T_{BD}$ as proven by equation 4, where $N$ is the number of brokers on the network. For every broker on the network, the master broker sends a TCP connection message and waits for TCP connection acknowledgement for a period of time called *TimeOut*. Generally, if the broker is active, the master broker will receive the acknowledgement message before the end of *TimeOut*.

$$T_{BD} = (\frac{N}{2})TimeOut + N(T_{TCP} + T_{TCPA}) \quad (4)$$

When the publisher changes the location of a topic from one broker to another, a subscriber requires a time $T_{change}$ that represents the time required to discover the new location of the topic and, and consequently, request the new broker to subscribe to that topic. $T_{change}$ is calculated as demonstrated through equation 5. Where $T_{TD}$ is defined in formula 3 while $T_{TTS}$ represents the delay needed by a subscriber to request the master broker about the location of the broker where the topics of interest exist and to send the subscription demand to the given broker. $T_{TTS}$ is the some the time needed to transmit the message exchanged between the master broker and the subscriber and the subscriber and the target broker, this delay is presented by formula 6.

$$T_{change} = T_{TD} + T_{TTS} \quad (5)$$

$$T_{TTS} = 2T_C + 2T_{CA} + 2T_S + 2T_{SA} + T_D \quad (6)$$

Now when a publisher detects an anomaly in the broker, he moves the location of the publication and the subscriber should be aware about brokers' anomaly, hence the subscriber should look about the new broker. For that purpose, the subscriber needs a time $T_{brokerChange}$ that represent the time needed of waiting an acknowledgment *TimeOut* from the broker and the time of the transparent topics subscriptions $T_{TTS}$ and the time of broker discovery $T_{BD}$ and the delay of the topics discovery $T_{TD}$ of each broker on the network, $N$ brokers exist in the network. $T_{brokerChange}$ is presented in equation 7.

$$T_{brokerChange} = TimeOut + T_{TTS} + T_{BD} + N.T_{TD} \quad (7)$$

### B. Evaluation Assumptions

Now we'll go through each of the assumptions that were used to assess the performance of the TD-MQTT technique one by one. These are our assumptions:

- The MQTT brokers are a mobile nodes.
- Messages are routed according to a routing protocol that is based on geographic locations. Instead than using routing tables, these protocols choose the shortest route path based on the nodes' location information [19]. Hence, nodes locations are known.
- When the publisher node discovers that a broker is more than two hops away, it will choose a broker within two hops. As a result, the broker's location is communicated with each message in order to determine if a publisher is close to or distant from the active broker.

### C. Results

The results of the scenario in which we proposed that the brokers' locations are variable are depicted in the figure 5. As shown in the same figure, we conclude that for the standard MQTT, the response time between the publisher and subscriber increases by enlarging the number of hops between the broker and the publisher, and as the subscriber and publisher have prior knowledge of the broker, the publisher can't change their broker and should connect continuously to that broker. As presented in the figure 5, despite that the broker are a mobile nodes, our approach (TD-MQTT) the response time between the publisher and the subscriber is much smaller than the response time of the standard MQTT, and this justified by the fact that the connection between the subscriber and the broker is done in a transparent way.

As introduced in the assumptions the publisher is always connected to a broker far of him of 2 hops, and as the broker nodes are mobile so the publisher should connect to another broker if the distance bigger than 2 hops and consequently change the location topics. Thus, the master broker should direct the subscriber to new topics location at their demands, for every new demand the master broker execute the topic discovery process to determine the address of the broker that contains the topics of interest. Figure 6 shows the delay needed to topic discovery process. In the EMMA [16] the subscribers and the publishers are connected to a separated broker. Every time, the broker move it is important to execute the broker discovery procedure provided by the authors of EMMA, that procedure search for the broker that respect the QoS between the subscriber and the broker. The figure 6 presents the difference between the TD-MQTT and EMMA approaches. The figure 6 clearly shows that the use of our approach is much more efficient since when a broker changes positions, the time needed to look up the new location of the broker where the topics of interest exist by our approach is much smaller than the time needed by the EMMA approach.

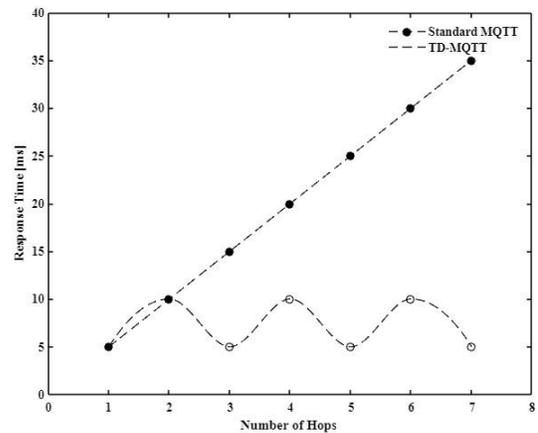

Fig. 5. Response Time between the Publisher and Subscriber by varying the Broker' Location

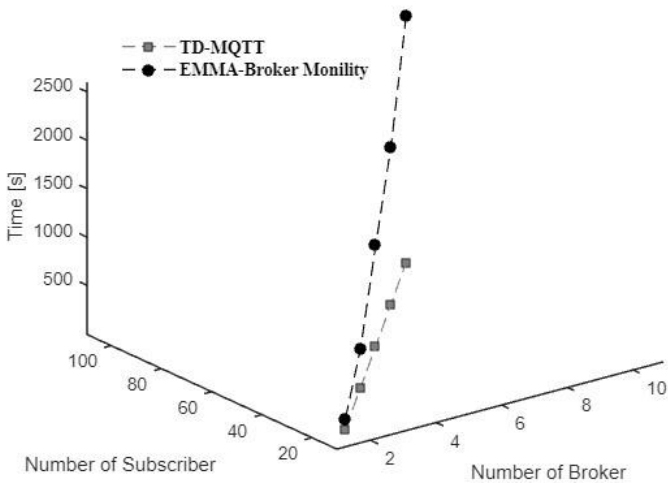

Fig. 6. Difference between TD-MQTT and EMMA-Broker Mobility

VI. CONCLUSION AND FUTURE WORKS

In this paper, we proposed a new version of MQTT called TD-MQTT. TD-MQTT allows the configuration of the network by providing a mechanism to transparently connect the subscribers to the active MQTT brokers. Instead of prior knowledge of the brokers to which the subscriber should connect, an external agent is responsible for the direction of the subscriber to the brokers that contain the topics of interest. This external agent is an MQTT broker, also known as a master broker. Hence, in a horizontal application that uses MQTT for communication, there is no need to know all the brokers' addresses. An industrialist shouldn't be aware of all the brokers' addresses. But it should only be aware of the address of the master broker that directs the subscribers to the appropriate brokers. The proposed approach allows the reconfiguration of the network in the case of the brokers' malfunctions, and consequently, the communication between publishers and subscribers does not know a stop time. We have discussed two scenarios to evaluate our approach. As the results of the evaluation show, we concluded that our approach is much better than the typical MQTT since the response time is much smaller. Different from the typical MQTT, our approach delivers a much smaller response time since every time the publishers and subscribers go away from the active broker, the network will be reconfigured (choose another broker).

Our approach would propose a mechanism to direct the subscribers to the active brokers. The direction mechanism is based on the topics' locations. In order to enhance that approach, we thought to ameliorate the direction mechanism by adding other features (e.g., memory and battery of the broker) to the brokers' selection. Also, we suggest simulating the proposed solution using an IoT-based case study.